\documentclass{aastex}          
\usepackage{spr-astr-addons}    

\catcode`\@=11
\newcommand{\gapprox}{\mathrel{\mathpalette\@versim>}}
\newcommand{\lapprox}{\mathrel{\mathpalette\@versim<}}
\newcommand{\propapprox}{\mathrel{\mathpalette\@versim\propto}}
\newcommand{\@versim}[2]
  {\lower3.1truept\vbox{\baselineskip0pt\lineskip0.5truept
\ialign{$\m@th#1\hfil##\hfil$\crcr#2\crcr\sim\crcr}}}
\catcode`\@=12

\begin{document}
%
\title{Particle acceleration in supernova-remnant shocks}

\shorttitle{Particle acceleration in SNRs}
\shortauthors{Reynolds}

\author{S.P. Reynolds\altaffilmark{}} 

\altaffiltext{1}{North Carolina State University}

\begin{abstract}
It has been known for over 50 years that the radio emission from shell
supernova remnants (SNRs) indicates the presence of electrons with
energies in the GeV range emitting synchrotron radiation.  The
discovery of nonthermal X-ray emission from supernova remnants is now
30 years old, and its interpretation as the extension of the radio
synchrotron spectrum requires electrons with energies of up to 100
TeV.  SNRs are now detected at GeV and TeV photon energies as well.
Strong suggestions of the presence of energetic ions exist, but
conclusive evidence remains elusive.  Several arguments suggest that
magnetic fields in SNRs are amplified by orders of magnitude from
their values in the ambient interstellar medium.  Supernova remnants
are thus an excellent laboratory in which to study processes taking
place in very high Mach-number shocks.  I review the observations of
high-energy emission from SNRs, and the theoretical framework in which
those observations are interpreted.
\end{abstract}


\section{Inferences from radio emission}

Remnants of historical supernovae had been known since Lundmark's
identification of the Crab Nebula with SN 1054 AD \citep{hubble28}.
The first radio source to be identified as a previously unknown
supernova remnant (SNR) was Cassiopeia A, proposed by Shklovskii in
1953, who also suggested that synchrotron radiation was the mechanism
for producing radio emission \citep{shklovskii53}, based on the
observed power-law spectrum $S_\nu \propto \nu^{-\alpha}$, with
$\alpha \sim 0.8$ for Cas A. Minkowski in 1957 confirmed the
identification from optical observations \citep{minkowski57}.  Thus
from the early days of radio astronomy, it was recognized that
energetic particles were present in SNRs.  Basic synchrotron physics
tells us that emission at frequency $\nu$ is primarily by electrons
with energy $E = 15 ( \nu({\rm GHz})/B(\mu{\rm
G}) )^{1/2}$ GeV, so radio emission at frequencies of a few
hundred MHz, typical at the time, immediately implied the presence of
electrons with Lorentz factors of $10^3 - 10^4$.

\begin{figure}
\includegraphics[width=2.5truein]{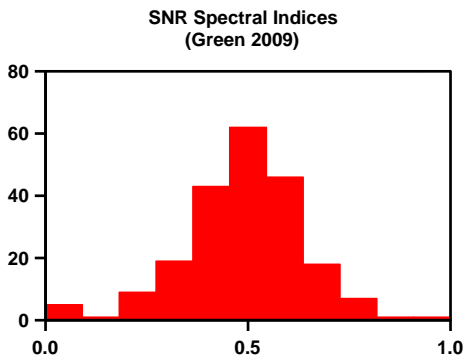}
\caption{Spectral-index distribution of SNRs in Green's catalogue
with adequately measured radio spectra, in bins of 0.1 in $\alpha.$
}
\label{spix}
\end{figure}

Radio remains the spectral region in which SNRs are most consistently
identified.  Green's famous catalogue of Galactic SNRs (Green 2009;
available online at http://www.mrao.cam.ac.uk/surveys/snrs/) lists 274
objects, essentially all with known radio properties.  The mean
spectral index is about 0.5, but with a significant spread of order
0.2 (Figure~\ref{spix}).  This spread is a significant problem for
theories of particle acceleration described below.  Also significant
is a tendency for the historical remnants (less than 2000 years old)
to have steeper indices ($\alpha \gapprox 0.6$), a trend continued by
radio supernovae, not represented here, which can have indices as
steep as 0.9 -- 1.0 \citep{weiler09}.  A few remnants such as Cas
A have very well-sampled radio spectra (see references in Green 2009);
Cas A's spectrum is well described by a single power-law with spectral
index of 0.77 between 100 MHz and 100 GHz.  However, most remnants are
represented by only a few data points with substantial error bars.
Other historical remnants have spectra with suggestions of concave-up
curvature \citep{reynolds92}, naturally explained by efficient shock
acceleration (see below).  

Magnetic field strengths are difficult to measure in SNRs; since the
intensity of synchrotron radiation from a power-law distribution of
electrons $N(E) = K E^{-s}$ electrons cm$^{-3}$ erg$^{-1}$ scales as
$K B^{1+\alpha}$, radio synchrotron fluxes only give roughly the
product of the energy densities in magnetic field and electrons.
X-ray evidence described below indicates that magnetic fields are
substantially amplified over typical interstellar values of a few
microGauss, but the only avenue for estimating energetics from
radio data is the use of minimum-energy equipartition arguments.
These arguments indicate that the minimum energy in electrons and
magnetic field in typical SNRs is far below the $\sim 10^{51}$ erg
explosion energies; SNRs are not efficient synchrotron radiators.
The equipartition magnetic fields so derived tend to be low, but as
there is no strong physical argument that equipartition should
hold, or even among which particles (should one include ions?),
the equipartition magnetic field strength is really just a proxy
for mean surface brightness.

Magnetic-field orientations can be usefully derived from radio
polarization directions.  A uniform synchrotron source with spectral
index $\alpha = 0.5$ has a polarized fraction of about 70\%, but very
few remnants show values above 40\%.  Young SNRs have much lower
values, typically of order 10\% -- 15\% (see references in Reynolds \&
Gilmore 1993), implying that their magnetic fields are primarily
disordered.  The ordered components, however, tend to be radial.  In
older remnants, magnetic-field orientations are typically confused,
but it is fairly common to see a tangential orientation, which one
would expect if a high-compression radiative shock compresses upstream
magnetic field, increasing the tangential component by a factor of the
compression ratio.

Even though radio observations of SNRs do not require a large fraction
of the SN energy, it is possible to argue that young SNRs require
acceleration of new electrons -- simply borrowing and compressing
relativistic electrons from the cosmic-ray population in the ISM is
inadequate both because of high observed surface brightnesses and
spectra much different from those of low-energy cosmic-ray electrons
\citep{reynolds08a}.

Radio observations of SNRs have left several unexplained puzzles, most
several decades old now.  What is responsible for spectral indices
flatter than 0.5?  (While a few pulsar-wind nebulae contaminate the
low-$\alpha$ end of the distribution of Figure 1, most of the remnants
with $\alpha < 0.5$ are shells.)  In a few cases, contamination with
flat-spectrum thermal emission may be responsible.  Steeper spectra
than 0.5 can be obtained with very low Mach-number shocks, but this
requires ${\cal M} \lapprox 10$ and is unlikely to be the case for
as many remnants as Figure~\ref{spix} requires.  Finally, the question
of the radial orientation of the ordered component of magnetic field
in young remnants remains unexplained, though the operation of
fluid instabilities at the contact interface between shocked ISM and
shocked ejecta is often invoked (e.g., Jun, Jones, \& Norman 1996).

\begin{figure}
\includegraphics[width=2truein]{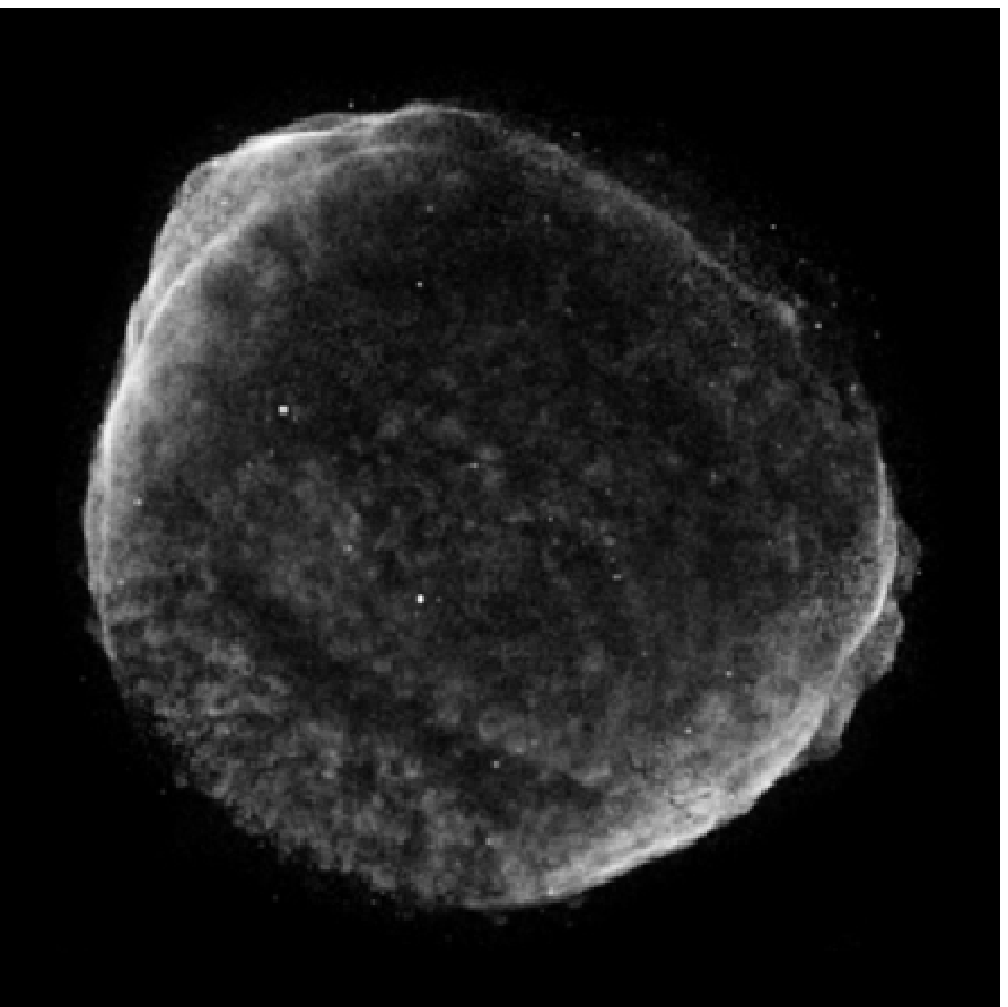}
\includegraphics[width=2truein]{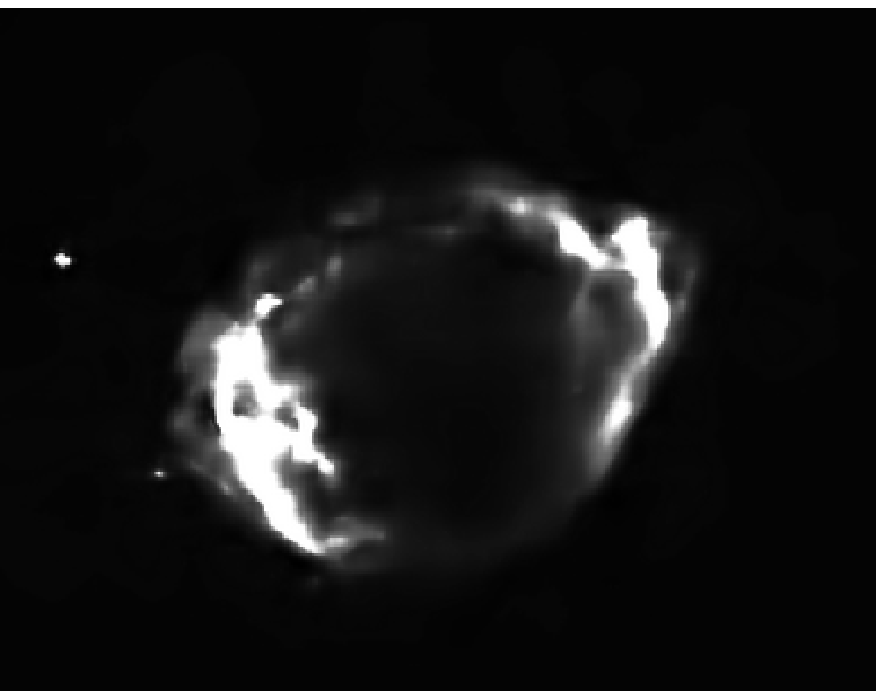}
\caption{{\sl Chandra} images of (top) SN 1006 (NASA/CfA) and (bottom)
G1.9+0.3 \citep{reynolds08b} between 1.5 and 7 keV, smoothed with
platelets as described in \cite{willett07}.}
\label{g1.9}
\end{figure}

\section{X-ray synchrotron emission}

The featureless X-ray spectrum of the remnant of SN 1006 above 1 keV
\citep{becker80,toor80} was first attributed to synchrotron radiation
from shock-accelerated electrons by \cite{reynolds81}.  This
explanation required the presence of electrons with TeV energies.
However, detection of oxygen lines from the spatially integrated
spectrum \citep{galas82} indicated the operation of thermal processes.
Only with the ASCA observation of thermal line emission from the
interior and a featureless continuum from the bright limbs (Koyama et
al.~1995) was it clear that a synchrotron component was called for,
though it represented the rolling off of the spectrum due to some
process limiting the maximum energy of electrons, rather than to a
straight extrapolation from radio \citep{reynolds96}.

How do we know the featureless continuum emission is synchrotron
radiation?  Nothing else works.  A nonthermal power-law electron
distribution with energies of order tens of keV will produce a
power-law bremsstrahlung spectrum with photon energies smaller by a
factor of a few.  However, those electrons would be just as efficient
at exciting atomic lines as electrons drawn from a Maxwellian
distribution.  A tiny corner of parameter space might be available for
a plasma either not yet ionized to the stages required for X-ray
lines, or on the other hand totally stripped
\citep[e.g.,][]{laming98}.  However, the careful study by
\cite{laming98} shows that in detail this cannot work for SN 1006.
Finally, inverse-Compton upscattering of any known source of photons
would produce a spectrum with the same slope as that of the
synchrotron emission produced by those electrons -- that is, the radio
slope ($\alpha \sim 0.6$), far too flat for the observations
\citep[$\alpha_x \sim 2.3$;][]{long03}. 
Furthermore, the synchrotron hypothesis makes a prediction: spectral
steepening to higher X-ray energies.  For SN 1006, this has been
confirmed by INTEGRAL \citep{kalemci06}.

There are now four known Galactic remnants whose soft X-ray spectrum
is dominated by synchrotron X-rays: in addition to SN 1006, G1.9+0.3
\citep{reynolds08b}; G347.3-0.5 \citep[also known as GX
J1713.7-3946][] {slane99}; and G266.2-1.2, ``Vela Jr.''
\citep{aschenbach98, slane01}. (See Reynolds 2008a for a more detailed
review of SNR X-ray and $\gamma$-ray emission.) Figure~\ref{g1.9}
shows the {\sl Chandra} images of SN 1006 and G1.9+0.3.  For both
objects, thermal lines have been detected from fainter regions of the
remnant.  The low-energy (central) emission in Fig.~\ref{g1.9} is
primarily ejecta emission dominated by oxygen.  Figure~\ref{g1.9spec}
shows the spectrum from the interior of G1.9+0.3, with clear emission
lines of Si, S, and Ar.

Synchrotron X-ray emission contributes to the spectrum of several more
Galactic remnants (reviewed in Reynolds 2008a).  In historical (or
quasi-historical) remnants RCW 86 (SN 185?), Tycho (SN 1572), Kepler
(SN 1604), and Cas A (SN $\sim 1680$), ``thin rims'' of featureless
X-ray emission lie at the edges of the remnants, and are presumed to
indicate the outer blast wave. Furthermore, all these remnants along
with SN 1006 have been reported to have hard X-ray continua in
integrated spectral (non-imaging) observations with the PCA instrument
on RXTE \citep{allen99}.  It appears that synchrotron X-ray emission
from the blast wave is a common feature in young SNRs, less than a few
thousand years old. The implications are striking: electrons are
present in these objects with energies $E = 72 (h\nu/1 \ {\rm
keV})^{1/2} (B /10\ \mu{\rm G})^{-1/2}$ TeV.  Now the electron
distributions are not straight power-laws all the way from
radio-emitting energies (ca.~10 GeV); rather, in all known cases, the
observed X-rays fall below that extrapolation
\citep{reynolds99}, indicating that some limitation on
electron energies is taking place below 100 TeV (perhaps far below,
for those older SNRs with no evidence for synchrotron X-ray emission).
As will be summarized below, shock acceleration may be limited by the
finite age (or size) of the remnant, particle escape above some energy
due to absence of scattering waves upstream, or (affecting electrons
only) radiative losses.

\section{Diffusive shock acceleration}

In standard diffusive shock acceleration \citep[amply reviewed
in][]{blandford87}, particles scatter from magnetic inhomogeneities
borne by converging fluids on either side of a shock wave, gaining
energy with each reflection.  Most particles disappear downstream
after each return, but a decreasing number remains for further cycles,
producing a power-law distribution (see Bell 1978 for a kinetic-theory
probabilistic argument).  A well-known result of this process, for
energetically unimportant test particles, is a power-law distribution
with index dependent only on the shock compression ratio: $N(E)
\propto E^{-s}$ with $s = (r + 2)/(r - 1)$ where $r \equiv
\rho_2/\rho_1$ is the shock compression ratio, equal to 4 for strong
shocks.  This result applies to extreme-relativistic particles ($E
\cong pc,$ $p$ the momentum and $c$ the speed of light).  The
prediction is thus $s \cong 2$ which, for electrons, implies a
synchrotron spectral index $\alpha = (s - 1)/2 = 0.5$, in tolerable
agreement with spectra observed from Galactic SNRs.

The maximum energy to which particles can be accelerated depends on
the limiting mechanism.  We assume a diffusion coefficient $\kappa$
which scales with particle energy.  This results if the particle mean
free path is a multiple $\eta$ of its gyroradius: $\lambda_{\rm mfp} =
\eta r_g = \eta E/eB$ for ultrarelativistic particles, since $r_g =
\gamma m c^2/eB$ (cgs units).  Then the ``Bohm limit'' in which the
mean free path is a gyroradius is $\eta = 1$.  For weak turbulence,
one expects $\eta \ge 1$, though this may not be a hard physical
limit.  For a remnant of age $t$ with shock speed $u_{\rm sh}$, with
surroundings containing MHD scattering waves only up to a wavelength
$\lambda_{\rm max}$, the maximum energies scale as
\begin{eqnarray}
E_{\rm max} ({\rm age}) &\propto & t \, u_{\rm sh}^2 \, B\, \eta^{-1} \\
E_{\rm max} ({\rm escape}) &\propto &\lambda_{\rm max}\, B \\
E_{\rm max} ({\rm loss}) &\propto & u_{\rm sh}\, B^{-1/2}\, \eta^{-1/2}
\end{eqnarray}
In all cases, for $u_{\rm sh} \gapprox 1000$ km s$^{-1}$ and ages
above a few hundred years, maximum energies of 10 -- 100 TeV are
easily obtainable.  The accelerated-particle spectra should show
exponential cutoffs with these fiducial energies.  For synchrotron
emission, the dependence of the peak frequency 
$\nu_m$ on electron energy of $\nu_m \propto E^2$ means that the
synchrotron spectrum will then drop roughly as $\nu^{-\sqrt{(E/E_{\rm
max})}}$, that is, considerably slower than exponential, and hardly
differing from a power-law in the bandpass of most X-ray observatories
($\sim 0.3 - 10$ keV).  

The diffusion coefficient may be anisotropic; in particular, diffusion
along and across magnetic-field lines is likely to take place at
different rates, with effects on the acceleration time $\tau$ to some
energy.  If the shock velocity makes an angle $\theta_{\rm Bn}$ with
the mean upstream magnetic field, we can parameterize this effect with
$R_J(\theta_{\rm Bn}, \eta, r) \equiv \tau(\theta_{\rm
Bn})/\tau(\theta_{\rm Bn} = 0)$.  We scale to typical values for young
SNRs: $u_{8.5} \equiv u_{\rm sh}/3000$ km s$^{-1}$; $t_3 \equiv t/1000$
yr; $B_{10} \equiv B/10 \ \mu{\rm G}$; and $\lambda_{17} \equiv
\lambda_{\rm max}/10^{17}$ cm.  The frequencies at which electrons
with energy $E_{\rm max}$ emit their peak power for each cutoff
mechanism are then
\begin{eqnarray}
h\nu_{\rm roll} ({\rm age}) &\sim& 0.4\, u_{8.5}^4\, t_3^2\, B_{10}^3\, (\eta R_J)^{-2}\ {\rm keV} \\
h\nu_{\rm roll} ({\rm esc}) &\sim& 2 \,B_{10}^3\, \lambda_{17}^2 \ {\rm keV} \\
h\nu_{\rm roll} ({\rm loss}) &\sim& 2\, u_{8.5}^2\, (\eta R_J)^{-1}\ {\rm keV}.
\end{eqnarray}
Of course, in a given object, the lowest value of $E_{\rm max}$ will
be the operative value.  Thus if one can determine the mechanism
causing the spectral cutoff, its value constrains considerably more
physical parameters than the simple observation of a radio power-law
spectrum.

\section{Radiation from GeV to TeV photon energies}

Populations of relativistic ions and electrons can produce observable
continuum radiation through four mechanisms, one hadronic and three
leptonic (reviewed in Reynolds 2008a).  The hadronic mechanism is the
inelastic scattering of cosmic-ray protons on thermal nuclei,
producing pions.  The charged pions decay to electrons and positrons,
making a (probably) negligible contribution to the relativistic lepton
pool, but the $\pi^0$'s decay to gamma rays of comparable energy
$E_\gamma ({\rm min}) = m_\pi c^2/2 \sim 70$ MeV.  The spectrum of
emitted photons should be that of the ions that produce them.  The
three leptonic processes are synchrotron emission, described above, as
well as nonthermal bremsstrahlung, with the same spectrum as that of
the nonthermal electrons, and inverse-Compton (IC) upscattering of
photons from any significant ambient radiation.  In practice, this is
likely to be primarily the cosmic microwave background (CMB), though
in some cases, IC from UV-optical-IR photons may be
competitive.  The spectrum will be the same as that of the synchrotron
emission from whatever population of particles is responsible, that
is, considerably harder in the keV -- TeV range than that of the other
processes.  While the synchrotron process is clearly operating, it is
not clear which of the other processes might be responsible for
emission from any particular object.  The best evidence for ion
acceleration is the spectral feature resulting from the minimum photon
energy from a created pion nearly at rest, about 70 MeV.  Detailed
calculations \citep[e.g.,][]{baring99} show that this feature may not
be highly distinct in a real object.

Several shell (i.e., not containing a pulsar) SNRs have been detected
in TeV gamma rays, using air-\v{C}erenkov detectors such as the
High-Energy Stereoscopic System in Namibia.  These include G347.3-0.5,
Vela Jr., RCW 86, and SN 1006 \cite[see references in][]{reynolds08a}.
The spectra are steep, with photon indices $\Gamma \sim 2$ ($F_\gamma
\propto (h\nu)^-\Gamma$); for G347.3-0.5, the spectrum is observed to
steepen above 1 TeV.  Elaborate models for these
four cases have been constructed \citep[e.g.,][]{berezhko06}.  The TeV
emission in these models can be due either to IC from the CMB, or to
$\pi^0$ decay.  Both classes of model have difficulty.  The former
require low filling factors of magnetic field, and imply inefficient
shock acceleration, while the latter suffer from severe limits on
thermal gas from X-ray observations, implying insufficient targets for
the relativistic protons \citep{ellison10}.  For complex objects such
as G347.3-0.5 and Vela Jr., simple one-zone models may be inadequate,
though there is as yet no clear path to consistent models.

\begin{figure}
\includegraphics[width=2truein]{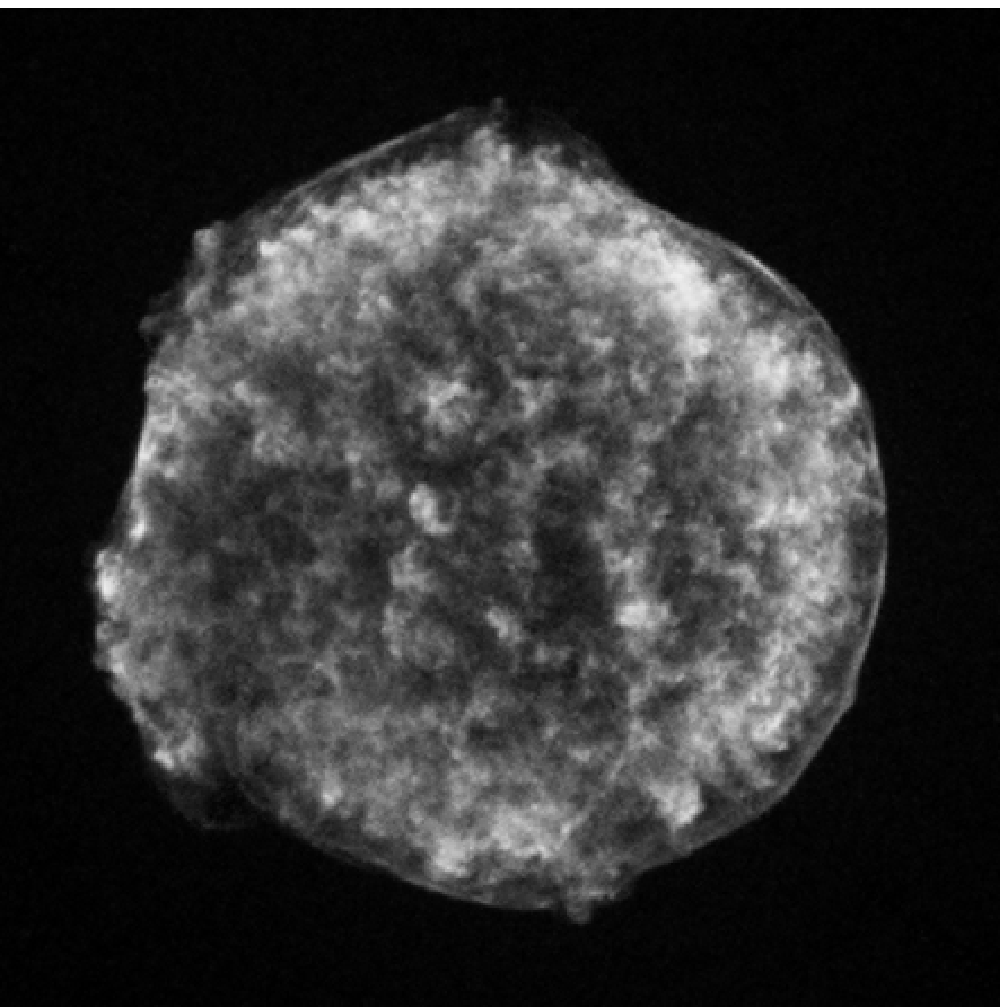}
\includegraphics[width=2truein]{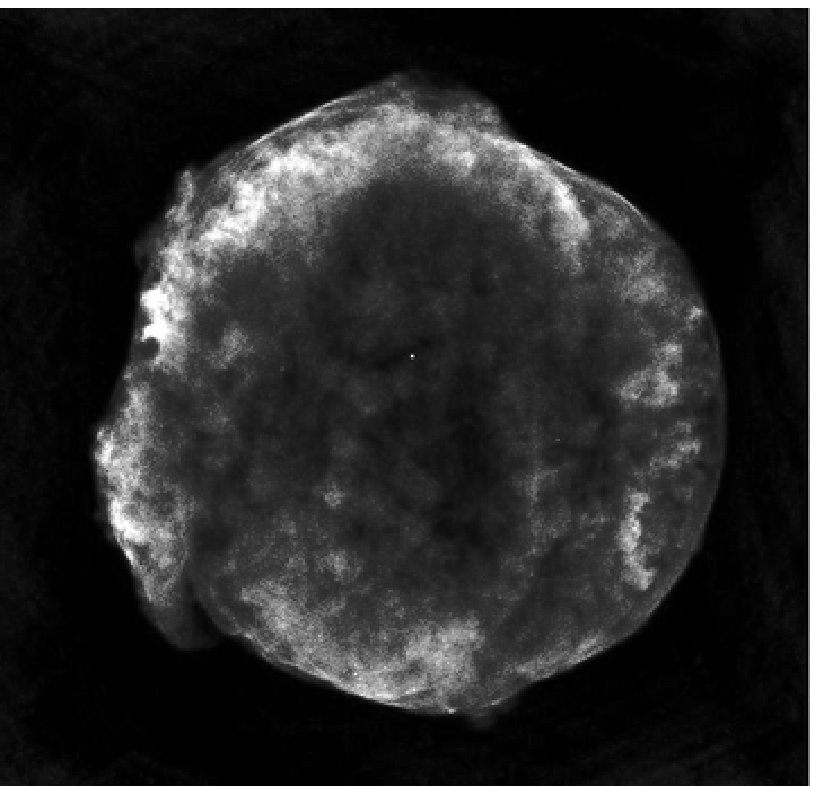}
\caption{Top:  {\sl Chandra} image of Tycho's SNR (NASA/CfA).
Below: VLA image at 1.4 GHz \citep{reynoso97}.  Note the thin
rim in both bands at the NE edge (upper left).}
\label{tycho}
\end{figure}

\section{Magnetic-field amplification}

An important result first suggested many years ago, but only recently
put on a firmer observational foundation, is the increase in
magnetic-field strength in SNRs over a simple factor of the
compression ratio expected in highly ionized gases.  The possibility
of much stronger shock amplification of magnetic fields was first
proposed for heliospheric shocks by Chevalier (1977), while the models
of \cite{reynolds81} demanded substantial amplification.  Early
gamma-ray upper limits from Cas A bounded the electron population from
above (from the inferred absence of bremsstrahlung), bounding the
magnetic-field strength from below based on observed radio synchrotron
fluxes.  \cite{cowsik80} used this argument to deduce a minimum
magnetic field strength of about 1 mG.  The slight concave curvature
observed in radio spectra of SNRs was explained by \cite{reynolds92}
as due to efficient shock acceleration modifying the shock structure,
but values of magnetic field were also implied of 100 $\mu$G and more
-- far higher than a few times the typical interstellar magnetic field
of 3 -- 5 $\mu$G.

High-resolution X-ray images from {\sl Chandra} show that the ``thin
rims'' present in most historical shell SNRs are in fact very thin --
so thin that an unusual depletion (beyond simple post-shock expansion)
of either relativistic electrons or magnetic field is required to
explain the sudden disappearance of synchrotron emissivity downstream
\citep{bamba03,vink03}.  If synchrotron losses are depleting the
electrons, we infer \citep{parizot06}
\begin{equation}
B > 200 u_8^{2/3} (w/0.01 \ {\rm pc})^{-2/3} \mu{\rm G}
\end{equation}
where the shock speed $u_8 \equiv u/1000$ km s$^{-1}$ and the filament
width is $w$.  However, if the amplified magnetic field is in waves of
some kind, these waves might damp, providing an alternative mechanism
for the rims \citep{pohl05}.  Figure~\ref{tycho} shows a typical rim
-- but it is present in radio as well, where synchrotron losses cannot
possibly be a factor.  It is possible that both mechanisms are
required to explain some rims.

An additional argument for the amplification of magnetic field comes
from the variations seen in periods of a few years in X-ray features
in G347.3-0.5 \citep{uchiyama07} and Cas A \citep{patnaude07}.  Both
brightening and fading are seen.  If the relevant timescales are those
of particle acceleration and synchrotron losses, in either case one
obtains values of the magnetic field of order 1 mG.  However, if the
post-shock magnetic field is highly turbulent, one expects such
``twinkling'' due to fluctuations even if the mean magnetic field is
lower \citep{bykov08}.

\section{The youngest Galactic remnant G1.9+0.3}

\begin{figure}
\includegraphics[width=2truein]{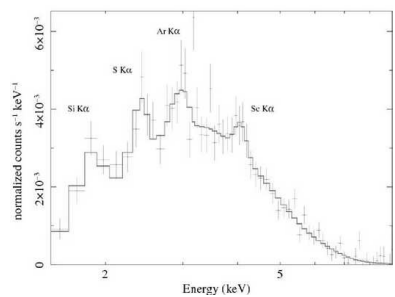}
\caption{Spectrum from the central region of G1.9+0.3 (Borkowski et al., in
preparation).}
\label{g1.9spec}
\end{figure}

The discovery of a remnant only about 100 yr old, from expansion
between radio observations made in 1985 and X-ray {\sl Chandra}
observations in 2007 \citep{reynolds08b}, opens a new window on the
very early development of a SNR, and of the physics of unprecedentedly
fast SNR shocks.  At an assumed distance of the Galactic Center, 8.5
kpc (the very high absorbing column, $N_H \sim 6 \times 10^{22}$
cm$^{-2}$, means the distance cannot be much less), the mean expansion
speed is about 14,000 km s$^{-1}$.  The integrated synchrotron
spectrum has a rolloff frequency of $h\nu = 2.2$ keV, one of the
highest ever measured.  A recent longer observation has revealed the
presence of spectral lines from the radio-bright N rim and interior
(Figure~\ref{g1.9spec}).

The figure shows clear lines of helium-like states of Si, S, and Ar
from the central region. Ca, while visible from the N rim, is absent
in the center -- but a line at 4.1 keV, which we identify as due to
$^{44}$Sc, is clearly present.  $^{44}$Sc is produced by electron
capture in $^{44}$Ti, a radioactive element with mean life of 89 yr.
The inner-shell vacancy is filled by the emission of a 4.1 keV photon.
This is the first firm detection of this element in a SNR.  The line
strength we measure implies an initial mass of $^{44}$Ti of $(1 - 3)
\times 10^{-5}\ M_\odot$ for an age of 100 -- 140 yr, within the
predicted range for either core-collapse or Type Ia supernovae, though
earlier spherically symmetric Type Ia models predict lower $^{44}$Ti
masses.  Since the $^{44}$Sc need not be ionized to produce the 4.1 keV
X-rays, we are sensitive to both shocked and unshocked $^{44}$Sc.  A
longer observation can allow the determination of the spatial
distribution and kinematics of $^{44}$Ti, of considerable interest to
SN modelers.

\section{Status report}


\begin{enumerate}

\item X-ray observations show that SNR shocks routinely exhibit
synchrotron radiation from electrons up to 100 TeV in energy.

\item Standard diffusive shock acceleration can easily account for
these energies for a range of conditions.  Little, however, can be
predicted.  Quantities of interest include efficiencies,
obliquity-dependence (i.e., $\theta_{\rm Bn}$-dependence), or e/p
ratio.

\item Emission at TeV energies seen from several young SNRs may be
leptonic (IC/CMB) or hadronic ($\pi^0$-decay).  Both types of models
have serious problems.  In particular, while many indirect lines of
evidence suggest that these shocks efficiently accelerate ions, direct
observational confirmation is still lacking.

\item Both integrated fluxes and small-scale structure in X-rays
indicate that magnetic fields in SNRs are higher by 10 -- 100 than
expected from simple compression of ISM fields.  This may be due to a
cosmic-ray driven instability \citep{bell04}.  This
instability requires high-efficiency ion acceleration.

\item Future needs: Observations at GeV energies (and, we hope, down
to 100 MeV or below) with the {\sl Fermi} Gamma-Ray Space Telescope
will be important in constraining models.  Theoretical work to predict
how particles are injected into the shock-acceleration process as well
as the items enumerated above is essential.  Radio observations
provide information which has still not been used by modelers to
maximum effect.  Finally, G1.9+0.3 is evolving so quickly that its
monitoring with time, and the accumulation of better statistics for
spectroscopy, may add substantially to our understanding of shock
acceleration in young SNRs.

\end{enumerate}

\acknowledgments

I gratefully acknowledge support from NASA and NSF for supernova-remnant
research.


\begin{thebibliography}{}

\bibitem[Allen, Gotthelf, \& Petre(1999)]{allen99}
Allen, G.E., Gotthelf, E.V., \& Petre, R. 1999, Proc.26th ICRC, 3, 480

\bibitem[Aschenbach(1998)]{aschenbach98}
Aschenbach, B. 1998, Nature, 396, 141

\bibitem[Bamba et al.(2003)]{bamba03}
Bamba, A., et al. 2003, ApJ, 589, 827

\bibitem[Baring et al.(1999)]{baring99}
Baring, M.G., et al. 1999, ApJ, 513, 311

\bibitem[Becker et al.(1980)]{becker80}
Becker, R.H., et al. 1980, ApJ, 240, L33

\bibitem[Bell(1978)]{bell78}
Bell, A.R. 1978, MNRAS, 182, 147

\bibitem[Bell(2004)]{bell04}
Bell, A.R. 2004, MNRAS, 353, 550

\bibitem[Berezhko \& V\"olk(2006)]{berezhko06}
Berezhko, E.G., \& V\"olk, H. 2006, A\&A, 451, 981

\bibitem[Blandford \& Eichler(1987)]{blandford87}
Blandford, R.D., \& Eichler, D. 1987, Phys.Rep., 154, 1

\bibitem[Bykov et al.(2008)]{bykov08}
Bykov, A.M., et al. 2008, ApJ, 689, L133

\bibitem[Chevalier(1977)]{chevalier1977}
Chevalier, R.A. 1977, Nature, 266, 701

\bibitem[Cowsik \& Sarkar(1980)]{cowsik80}
Cowsik, R., \& Sarkar, S. 1980, MNRAS, 191, 855

\bibitem[Ellison et al.(2010)]{ellison10}
Ellison, D.C., et al. 2010, ApJ, 712, 287

\bibitem[Galas et al.(1982)]{galas82}
Galas, C.M.F., Venkatesan, D., \& Garmire, G. 
1982, Ap.Lett., 22, 103

\bibitem[Green(2009)]{green09} Green, D.A. 2009, ``A Catalogue of
Galactic Supernova Remnants (2009 March version)'', Astrophysics
Group, Cavendish Laboratory, Cambridge, UK

\bibitem[Hubble(1928)]{hubble28}
Hubble, E. 1928, Astr.Soc.Pacific Leaflet \#14

\bibitem[Jun, Jones, \& Norman(1996)]{jun96}
Jun, B.-I., Jones, T.W., \& Norman, M.L.
1996, ApJ, 468, L59

\bibitem[Kalemci et al.(2006)]{kalemci06}
Kalemci, E., et al. 2006, ApJ, 644, 274

\bibitem[Koyama et al.(1995)]{koyama95} 
Koyama, K., et al. 1995,
Nature, 378, 255

\bibitem[Laming(1998)]{laming98}
Laming, J.M. 1998, ApJ, 499, 309

\bibitem[Long et al.(2003)]{long03}
Long, K.S., et al. 2003, ApJ, 586, 1162

\bibitem[Minkowski(1957)]{minkowski57}
Minkowski, R. 1957, Proc.IAU Symp.\#4, ed. H.C.van de Hulst
(Cambridge: Cambridge U. Press), 107

\bibitem[Parizot et al.(2006)]{parizot06}
Parizot, E., et al. 2006, A\&A, 453, 387

\bibitem[Patnaude \& Fesen(2007)]{patnaude07}
Patnaude, D.J., \& Fesen, R.A. 2007, AJ, 133, 147

\bibitem[Pohl et al.(2005)]{pohl05}
Pohl, M., et al. 2005, ApJ, 626, L101

\bibitem[Reynolds(1996)]{reynolds96}
Reynolds, S.P. 1996, ApJ, 459, L13

\bibitem[Reynolds(2008a)]{reynolds08a}
Reynolds, S.P. 2008a, ARA\&A, 46, 89

\bibitem[Reynolds et al.(2008b)]{reynolds08b}
Reynolds, S.P. et al. 2008b, ApJ, 680, L41

\bibitem[Reynolds \& Chevalier(1981)]{reynolds81}
Reynolds, S.P., \& Chevalier, R.A. 1981, ApJ, 245, 912

\bibitem[Reynolds \& Ellison(1992)]{reynolds92}
Reynolds, S.P., \& Ellison, D.C. 1992, ApJ, 399, L75

\bibitem[Reynolds \& Gilmore(1993)]{reynolds93}
Reynolds, S.P., \& Gilmore, D.M. 1993, AJ, 106, 272

\bibitem[Reynolds \& Keohane(1999)]{reynolds99}
Reynolds, S.P., \& Keohane, J.M. 1999, 525, 368

\bibitem[Reynoso et al.(1997)]{reynoso97}
Reynoso, E., et al.1997, ApJ, 491, 816

\bibitem[Shklovskii(1953)]{shklovskii53}
Shklovskii, I.S. 1953, Dokl.Akad.Nauk.SSSR 91, 475

\bibitem[Slane et al.(1999)]{slane99}
Slane, P.O., et al. 1999, ApJ, 525, 357

\bibitem[Slane et al.(2001)]{slane01}
Slane, P.O., et al. 2001, ApJ, 548, 814

\bibitem[Toor(1980)]{toor80}
Toor, A. 1980, A\&A, 85, 184

\bibitem[Uchiyama et al.(2007)]{uchiyama07}
Uchiyama, Y., et al. 2007, Nature, 449, 576

\bibitem[Vink \& Laming(2003)]{vink03}
Vink, J., \& Laming, J.M. 2003, ApJ, 584, 758

\bibitem[Weiler et al.(2009)]{weiler09}
Weiler, K.W., et al. 2009, AIP Conf.Proc., 1111, 440

\bibitem[Willett(2007)]{willett07} Willett, R. 2007, in Statistical
Challenges in Modern Astronomy IV, eds.~G.J.~Babu \& E.D.~Feigelson,
APS Conf.~Ser.~371, 247


\end{thebibliography}
\end{document}